\documentclass[12pt]{iopart}
\newcommand{\Jpsi}{$J/\psi$ }
\newcommand{\pT}{$p_T$ }
\newcommand{\xT}{$x_T$ }
\newcommand{\sNN}{$\sqrt{s_{\mathrm{NN}}}$ }
\newcommand{\pp}{$p+p$ }
\newcommand{\cucu}{Cu+Cu }
\newcommand{\raa}{$R_{AA}$ }
\usepackage{iopams}
\usepackage{graphicx}
\begin{document}

\title[\Jpsi  production at high \pT in \pp and $A+A$ collisions at STAR]
{\Jpsi production at high \pT in \pp and $A+A$ collisions at STAR}

\author{Zebo Tang\footnote{The author was supported in part by
the National Natural Science Foundation of China under Grant No.
10610285, 10610286, 10575101 and the Knowledge Innovation Project
of Chinese Academy of Sciences under Grant No. KJCX2-YW-A14.} (for
the STAR collaboration)}

\address{Dept. of Modern Physics, University of Science and Technology of China,
Hefei, Anhui, China, 230026; Brookhaven National Laboratory,
Upton, NY, 11973, USA} \ead{zbtang@mail.ustc.edu.cn}

\begin{abstract}
The preliminary results of \Jpsi spectra at high transverse
momentum ($5<p_T<14$ GeV/c) in \pp and \cucu collisions at \sNN =
200 GeV are reported. The nuclear modification factor is measured
to be $0.9\pm0.2$ at $p_T>5$ GeV/c. The correlations between \Jpsi
and charged hadrons are also studied in \pp collisions to
understand the \Jpsi production mechanism at high $p_T$.
\end{abstract}


\section{Introduction}
\Jpsi dissociation from color-screening of Quantum Chromodynamics
(QCD) in a Quark-Gluon Plasma (QGP) is one of the major signatures
of QCD de-confinement in relativistic heavy-ion collisions.
Recent calculations which assuming the AdS/CFT duality is valid
for QCD expect that a heavy fermion pair bound state (an analog of
quarkonium in QCD) will have an effective dissociation temperature
decreasing with $p_T$ \cite{adscft}.
This requires a measurement of \Jpsi extending to $p_T>5$ GeV/c
where the effective \Jpsi dissociation temperature is expected to
decrease to the temperature reached at RHIC ($\sim$ 1.5 $T_c$) .
In this paper, we report the \Jpsi spectra at high transverse
momentum ($5<p_T<14$ GeV/c) in \pp and \cucu collisions at \sNN =
200 GeV. In addition, we performed an analysis of $J/\psi$-hadron
correlations in \pp collisions to understand the \Jpsi production
mechanism at high $p_T$. The technique is similar to that used by
UA1 \cite{UA1} and dihadron correlations analyzed by STAR
\cite{fuqiang}.

UA1 simulated $J/\psi$-hadron correlation and found two cases:
When a \Jpsi originated from $\chi_c$ there as no visible
near-side correlation, whereas $J/\psi$'s originating from $B$
meson decays showed a strong near-side correlation. The large
acceptance of STAR Time Projection Chamber (TPC)
\cite{stardetector} and the Barrel Electromagnetic Calorimeter
(BEMC) \cite{stardetector} covering $|\eta|<1$ are very well
suited for such analyzes.

\section{Data analysis and Results}

 At STAR, both the TPC and BEMC can provide electron
identification \cite{stardetector}. At high $p_T$, the BEMC is
very powerful for electron identification and can also be used to
set up a fast trigger to enrich the electron sample. At moderate
$p_T$, the TPC can identify electrons efficiently. In this
analysis, the high \pT \Jpsi was reconstructed through the
dielectron channel, one electron at high \pT identified by combing
the BEMC and TPC and the other electron at lower \pT identified by
the TPC only. We used the BEMC triggered data in \pp and \cucu
collisions at \sNN = 200 GeV. The integrated luminosity is $\sim$
2.8 (\textit{11.3}) $pb^{-1}$ for \pp collisions collected in year
2005 (\textit{2006}) with transverse energy threshold $E_T>$ 3.5
(\textit{5.4}) GeV, and $\sim$ 860 $\mu b^{-1}$ for \cucu
collisions collected in year 2005 with $E_T>$ 3.75 GeV.

\begin{figure}[th]
\begin{minipage}[c]{0.5\textwidth}
\centering
\includegraphics[width=0.75\textwidth]{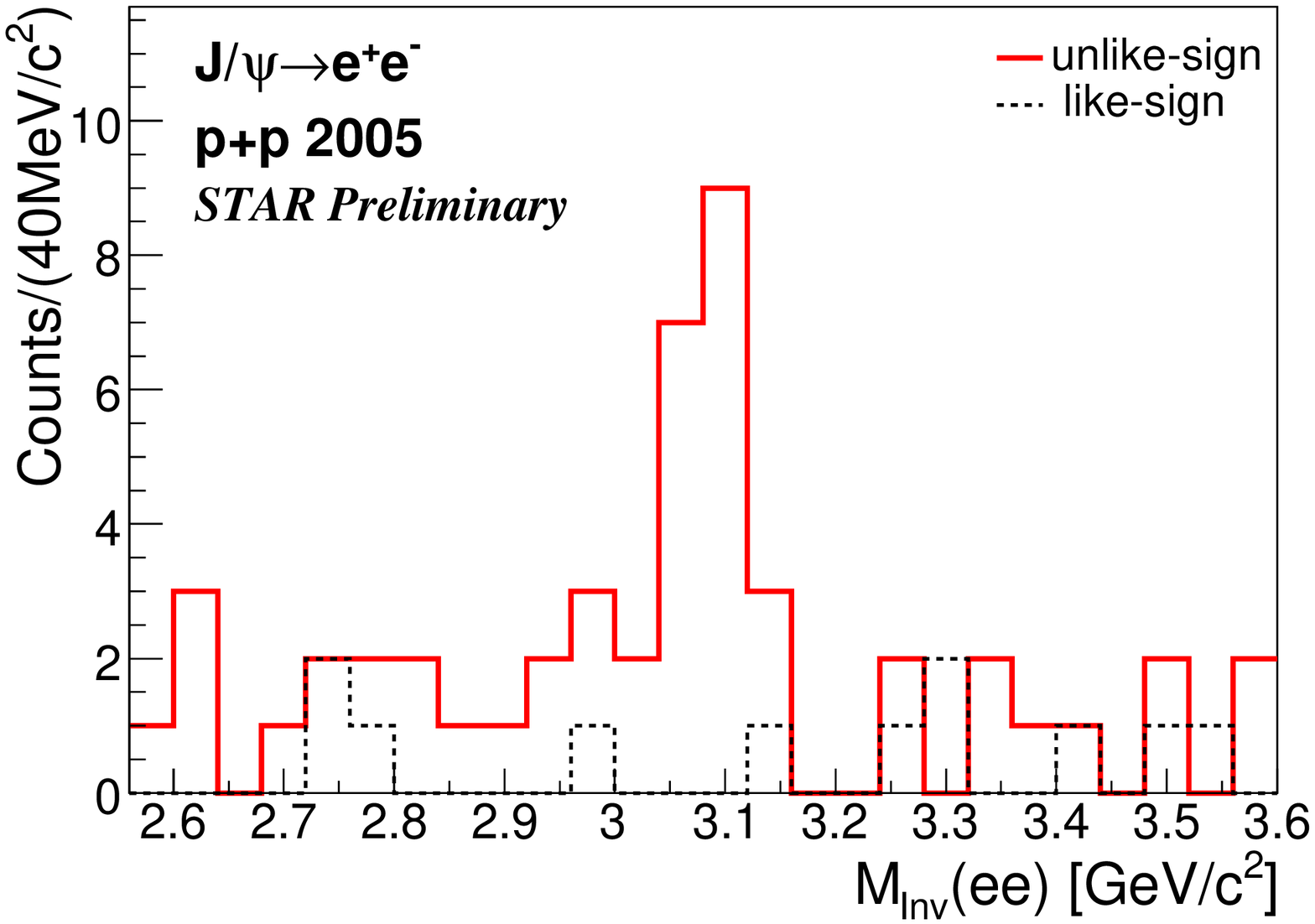}
\end{minipage}
\begin{minipage}[c]{0.5\textwidth}
\centering
\includegraphics[width=0.75\textwidth]{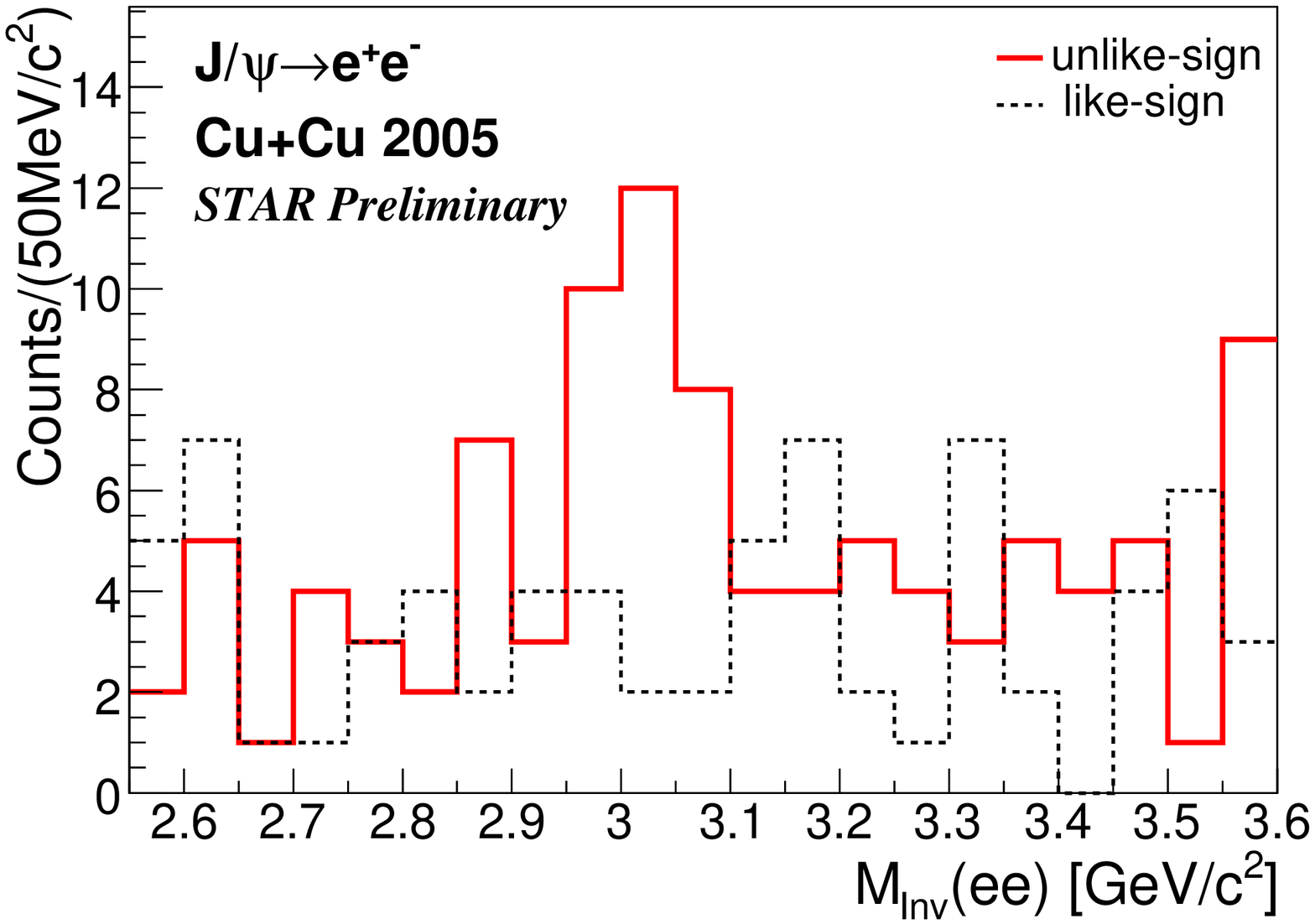}
\end{minipage}
\caption{The dielectron invariant mass distributions in \pp
(\textit{left}) and \cucu (\textit{right}) collisions at \sNN= 200
GeV.
} \label{invmass}
\end{figure}

\begin{figure}[th]
\begin{minipage}[c]{0.33\textwidth}\centering\mbox{
\includegraphics[width=0.98\textwidth,height=0.98\textwidth]{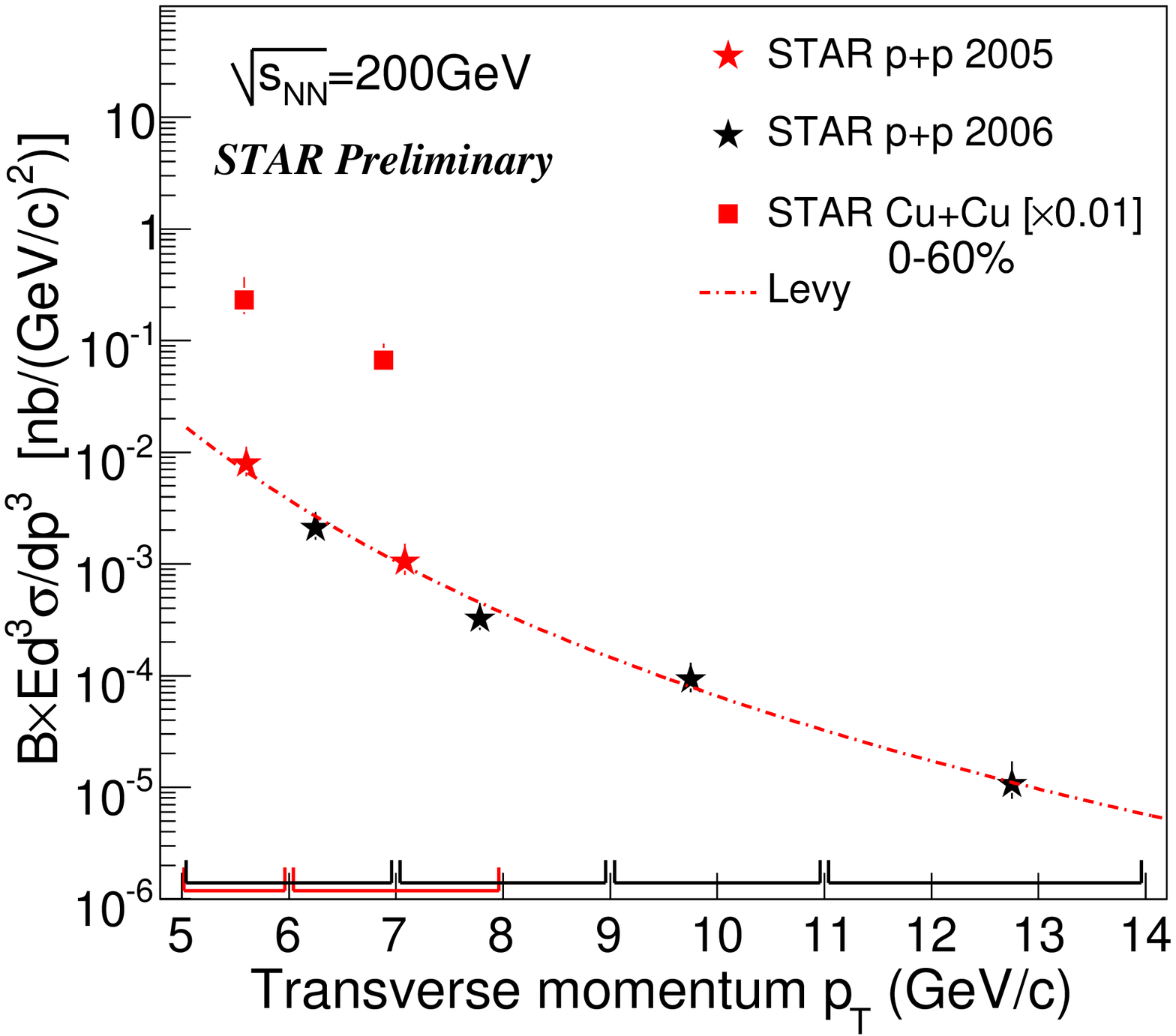}}
\end{minipage}
\begin{minipage}[c]{0.33\textwidth}\centering\mbox{
\includegraphics[width=0.98\textwidth,height=0.98\textwidth]{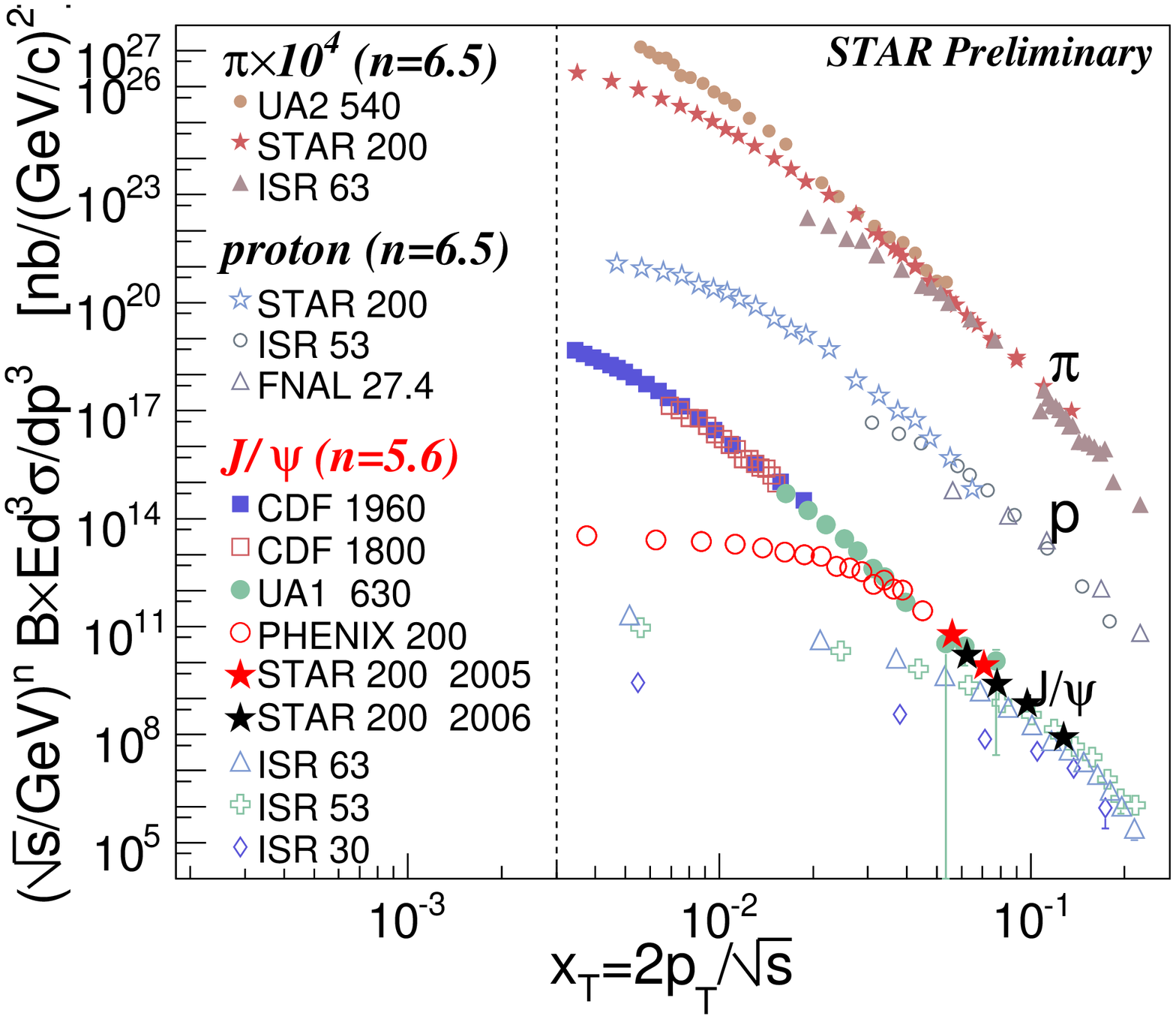}}
\end{minipage}
\begin{minipage}[c] {0.33\textwidth}\centering\mbox{
\includegraphics[width=0.98\textwidth,height=0.98\textwidth]{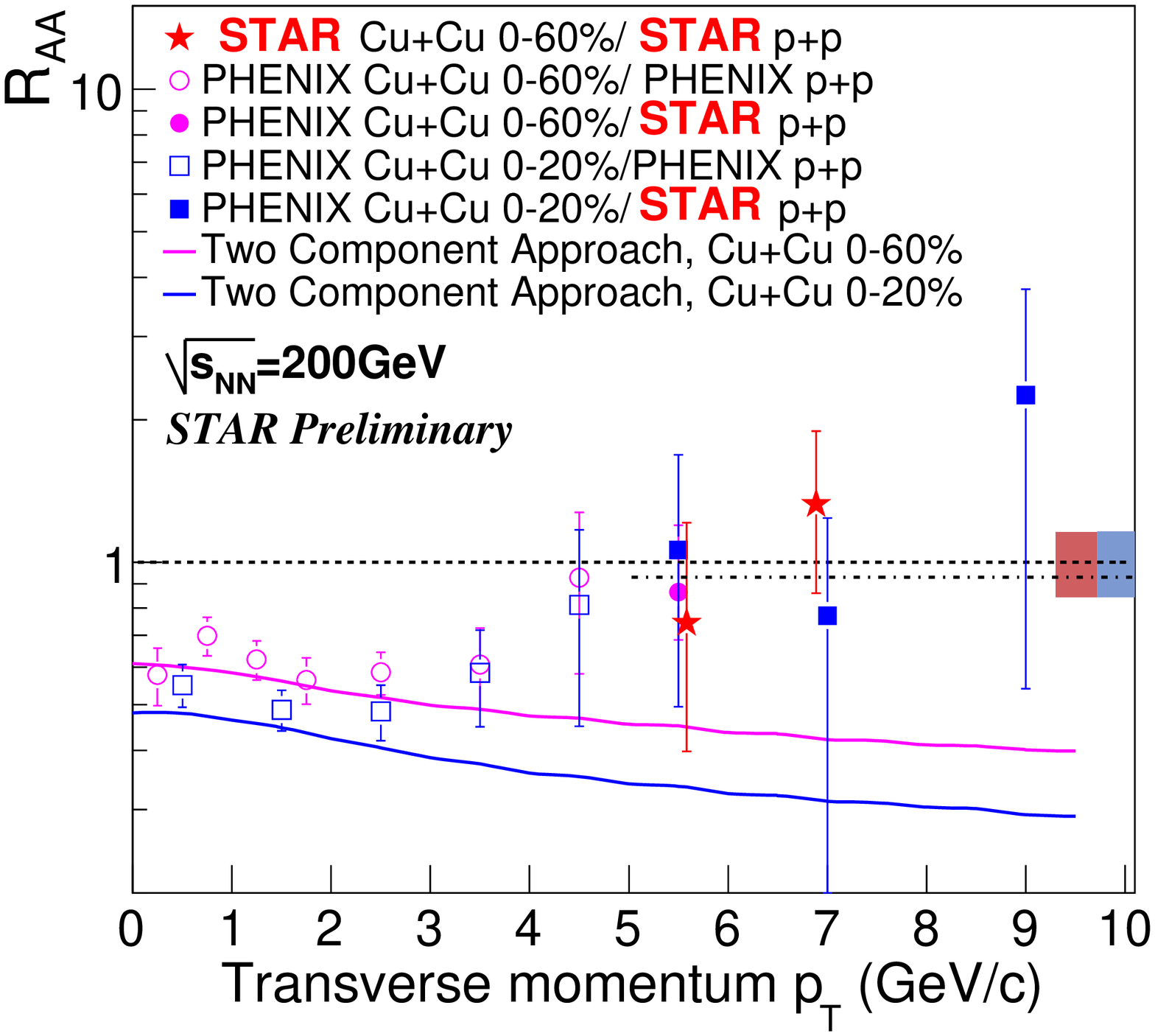}}
\end{minipage}
\caption{Left: \Jpsi invariant cross section
as a function of \pT in \pp and \cucu collisions at \sNN = 200
GeV. Errors shown are statistical only. Middle: \xT scaling of
pions, protons and $J/\psi$s. The data from other measurements can
be found in references
\cite{CDF,FNAL,ISR,phenixpp,scalingpi,UA1,UA2}. Right: \Jpsi
$R_{AA}$ as a function of $p_T$. The dot-dashed line represents
the fit by constant to all the data points at $5<p_T<10$ GeV/c.
The boxes on the right show the normalization uncertainty.}
\label{xsection}
\end{figure}

Figure \ref{invmass} shows the high \pT \Jpsi signal in \pp
(\textit{left}) and \cucu (\textit{right}) collisions at \sNN =
200 GeV. The background is represented by the dashed lines from
like-sign technique. We applied a cut of \pT $>2.5-4$ GeV/c to the
EMC triggered electrons and the cut of \pT $>1.2-1.5$ GeV/c for
lower \pT electrons. This ensured clean \Jpsi identification. The
signal/background (S/B) ratio in the analysis is 22/2
(\textit{40/14}) in \pp collisions using year 2005 (\textit{2006})
data and 17/23 in \cucu collisions. The \pT coverage in \pp  and
\cucu collisions taken in year 2005 is $5<p_T<8$ GeV/c, while in
\pp collisions taken in year 2006, the \Jpsi \pT can reach 14
GeV/c due to higher recorded luminosity and full BEMC coverage.
The \Jpsi invariant cross section
$B_{ee}\times Ed^3\sigma/dp^3$, after efficiency correction, are
shown as symbols in Figure \ref{xsection} (\textit{left}).

The invariant cross section of inclusive pion and proton
production in high energy \pp collisions have been found to follow
the \xT scaling law:
$E\frac{d^3\sigma}{dp^3}=\frac{g(x_T)}{\sqrt{s}^n}$, where
$x_T=2p_T/\sqrt{s}$. The value of the power $n$ depends on the
quantum exchanged in the hard scattering and is related to the
number of point-like constituents taking an active role in parton
model. It reaches 8 in the case of a diquark scattering and
reaches 4 in more basic scattering processes (as in QED). Figure
\ref{xsection} (\textit{middle}) shows the \xT scaling of
$J/\psi$, pion and proton. The power $n$ was found to be $6.5 \pm
0.8$ for pion and proton \cite{scalingpi} and $5.6 \pm 0.2$ for
$J/\psi$, which indicates that the high \pT \Jpsi production
mechanism is closer to parton-parton scattering.

Figure \ref{xsection} (\textit{right}) shows the \Jpsi nuclear
modification factor \raa as a function of \pT in 0-20\% and 0-60\%
\cucu from PHENIX \cite{phenixcucu} and STAR measurements. \raa
tends to increase from low to high $p_T$, although the error bars
currently do not allow to draw strong conclusions. One can
nevertheless do a combined fit to all the high-pt data and find
that $R_{AA}=0.9\pm0.2$. This result is in contrast to the
expectation from AdS/CFT-based models \cite{adscft} and from the
Two-Component model \cite{tca} which predict a decreasing \raa
with increasing $p_T$. This result could indicate that other \Jpsi
production mechanisms such as virtual photons or formation time
\cite{csmraa} play a role at high $p_T$.

\begin{figure}[th]
\begin{minipage}[c]{0.5\textwidth}
\includegraphics[width=0.79\textwidth]{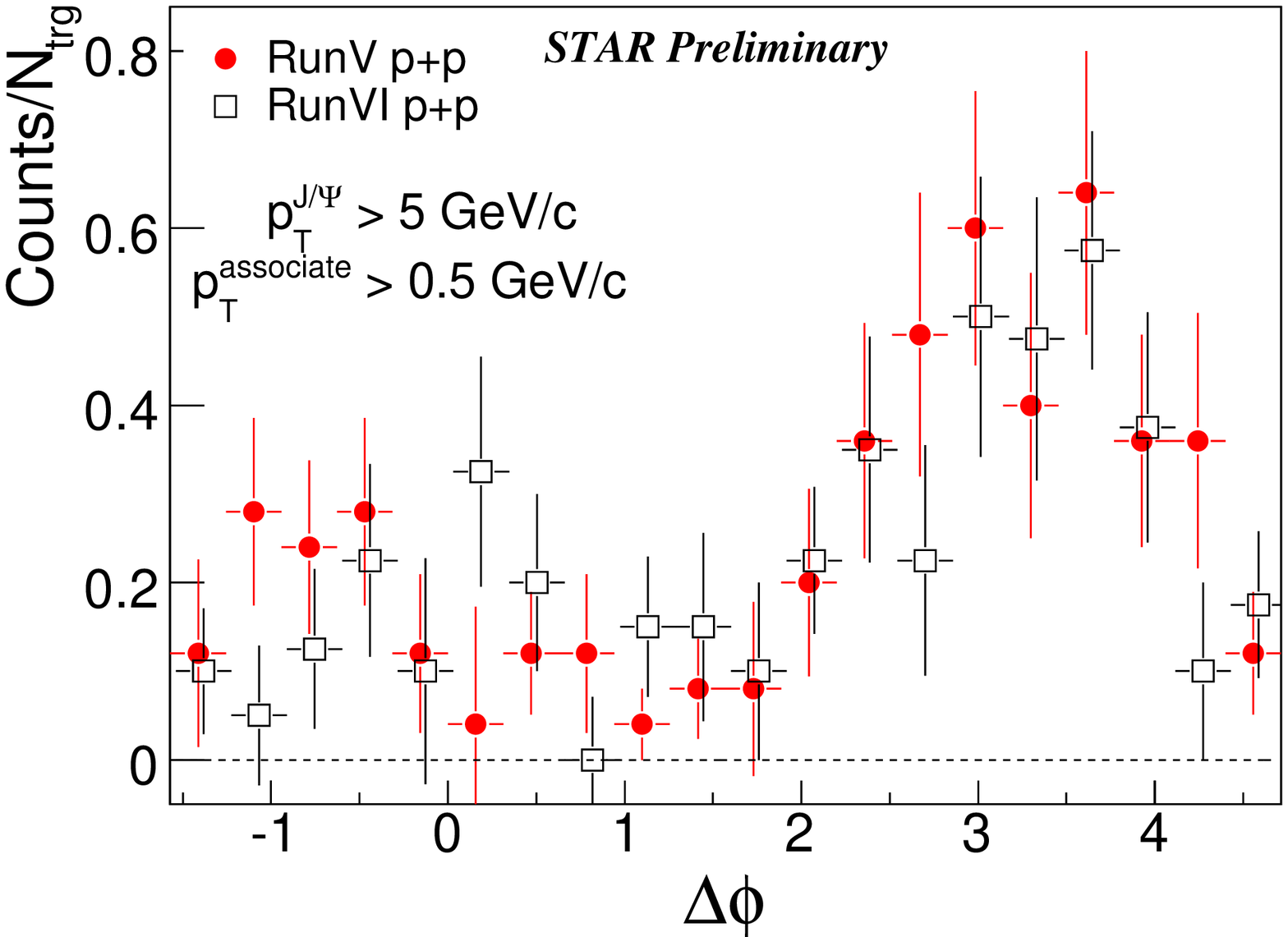}
\end{minipage}
\begin{minipage}[c]{0.5\textwidth}
\includegraphics[width=0.79\textwidth]{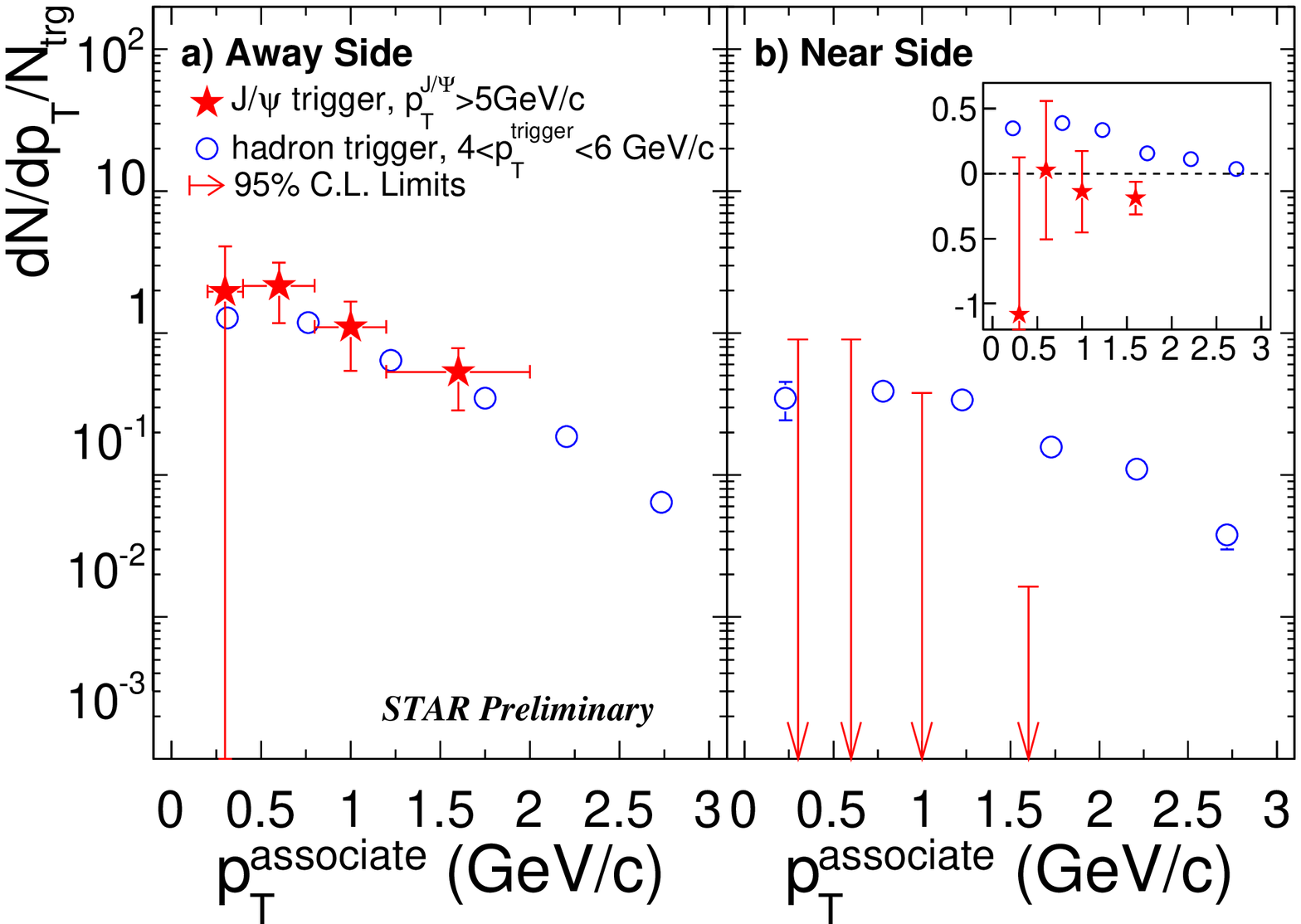}
\end{minipage}
\caption{Left: $J/\psi$-hadron correlations after background
subtraction in \pp collisions at \sNN= 200 GeV. Right: Associated
charged hadron \pT distribution on the near and away side with
respect to \Jpsi triggers and charged hadron triggers.}
\label{corr}
\end{figure}

With large S/B ratios, the $J/\psi$-hadron correlations were also
measured in \pp collisions. Figure \ref{corr} (\textit{left})
shows the azimuthal angle correlations between high \pT \Jpsi
($p_T>5$ GeV/c) and charged hadrons. No significant near side
correlations were observed, which is in contrast to the dihadron
correlation measurements \cite{fuqiang}. Since the Monte Carlo
simulation results show a strong near side correlation if the
\Jpsi is produced from $B$-meson decay, these results can be used
to constrain the $B$-meson contribution to \Jpsi production.
Figure \ref{corr} (\textit{right}) shows the associated charged
hadron \pT distribution on the near side and away side with
respect to \Jpsi triggers and charged hadron triggers.
On the away side, the yields of the associated charged hadrons
with respect to both kinds of triggers are consistent with each
other, which indicates that the hadrons on the away side of \Jpsi
triggers are from light quark or gluon fragmentation. On the near
side, the associated charged hadron yields with respect to \Jpsi
triggers are significantly lower than those with respect to
charged hadron triggers. This indicates that the $B$-meson is not
a dominant contributor to the inclusive high \pT $J/\psi$.

\section{Summary}

We reported the STAR preliminary results of \Jpsi spectra from 200
GeV \pp and \cucu collisions at high \pT ($5<p_T<14$ GeV/c) at
mid-rapidity through the dielectron channel. The high \pT \Jpsi
production was found to follow the \xT scaling with a beam energy
dependent factor $\sim$ \sNN$^{5.6\pm0.2}$. The \Jpsi nuclear
modification factor \raa in \cucu increases from low to high $p_T$
which challenges some models. The average of \raa at \pT$>$ 5
GeV/c is $0.9\pm0.2$, consistent with no \Jpsi suppression. It
implies that high \pT \Jpsi may be produced from virtual photon or
formed outside of the hot interaction region \cite{csmraa}. The
$J/\psi$-hadron correlations were also discussed. We observed an
absence of charged hadrons accompanying high \pT \Jpsi on the near
side which indicates that the $B$-meson is not a dominant
contributor to the inclusive high \pT $J/\psi$.

\end{document}